\documentstyle[12pt,twoside,fleqn,espcrc1,float,epsfig]{article}

\newcommand{\lsim}{%less than or approx. symbol
 \mathrel{\setbox0=\hbox{$<$}\raise0.6ex\copy0\kern-\wd0
 \lower0.65ex\hbox{$\sim$}}}

\newcommand{\gsim}{%less than or approx. symbol
 \mathrel{\setbox0=\hbox{$>$}\raise0.6ex\copy0\kern-\wd0
 \lower0.65ex\hbox{$\sim$}}}

\newcommand{\AmS}{{\protect\the\textfont2
  A\kern-.1667em\lower.5ex\hbox{M}\kern-.125emS}}

\title{QCD Aspects of Hadron Physics\thanks{Dedicated to Koichi Yazaki on the
occasion of his 60th birthday; invited talk presented at the KEK-Tanashi
Symposium on Physics of Hadrons and Nuclei, Tokyo, December 14--17, 1998. Work
supported in part by DFG and BMBF}}

\author{W. Weise\address{Physik-Department\\Technische Universit\"at
M\"unchen\\D-85747 Garching, Germany}}

\begin{document}
% typeset front matter
\maketitle

\begin{abstract}
 Several topics in hadron physics at different scales of resolution are
 discussed. First, deep-inelastic scattering from nucleons and nuclei is viewed
            in a light-cone coordinate space picture. Then the smooth
            transition from parton to hadron degrees of freedom is demonstrated
            by analysing generalized $Q^2$-dependent polarizabilities of the
            nucleon. Turning to low energy QCD we summarize recent developments
            related to the role of strange quarks in Chiral $SU(3)$ Dynamics, a
            non-perturbative coupled channel approach to hadronic
            processes. Finally we elaborate on a unification of QCD Sum Rules
            with aspects of spontaneous chiral symmetry breaking in the
            analysis of quark-antiquark excitations of the condensed QCD 
vacuum and in
            nuclear matter.
\end{abstract}

\section{\bf Introduction}
QCD has a comparatively "simple" basic Lagrangian but offers an enormous
variety of physics phenomena. They extend from asymptotically free quarks and
gluons at short space-time distances to complex hadronic
excitations at larger scales, built on a highly non-trivial vacuum which hosts
strong quark and gluon condensates. Given the genuine non-perturbative features
of QCD, namely confinement and spontaneous chiral symmetry breaking, the 
persistent
challenge is to identify and explore the active (effective) degrees of freedom
at each different scale. The present talk summarizes several topics and recent
results related to these issues. We start with an instructive little exercise,
elaborating quark and gluon distributions of nucleons and nuclei in coordinate
rather than momentum space. Such a description gives a simple geometrical
interpretation of the mechanisms at work in deep-inelastic lepton
scattering. Turning to lower $Q^2$ we outline the smooth transition from 
partonic
to hadronic degrees of freedom by the example of generalized ($Q^2$-dependent)
electromagnetic polarizabilities of the nucleon. Next we move to low-energy QCD
with strange quarks. We discuss recent developments and results of Chiral
$SU(3)$ Dynamics, our non-perturbative coupled channel approach based on the
chiral low-energy effective Lagrangian. In the final part of this survey, the 
QCD
Sum Rule method is unified with aspects of spontaneous chiral symmetry 
breaking in
an attempt to establish model-independent constraints for vector meson 
spectra, both in
vacuum and in nuclear matter.

\section{\bf Deep-inelastic lepton scattering in coordinate space} 
Nucleon structure functions are commonly analyzed in momentum space. In
coordinate space, quark and gluon distributions are defined as correlation
functions involving two field operators separated by a light-cone distance $y^+
= t + z = 2l$. In deep-inelastic scattering as viewed in the laboratory frame
with the target nucleon or nucleus at rest, the longitudinal distance $y^+ /2$
entering in the parton correlation function can be compared with length scales
characteristic of nucleons and nuclei and offers new insights into the nature
of parton distributions and their interpretation.

The quark and gluon distributions are expressed in terms of the squared
four-momentum transfer $Q^2$ and the Bjorken variable $x = Q^2 /2 M \nu$, where
$M$ is the nucleon mass and $\nu$ is the energy transfer in the lab frame. In
the Bjorken limit the dominant contributions to the structure functions at
small Bjorken-$x$ come from the light-like separations of order $y^+ \sim
1/Mx$. Consequently, large longitudinal distances $l = y^+/2 = (2 Mx)^{-1}$ are
important in the scattering process at small $x$.

The space-time pattern of deep-inelastic scattering is then as follows: the
virtual photon interacts with partons which propagate a distance $y^+$ along
the light cone. The characteristic lab frame correlation length $l$ is half of
that distance. These features are naturally implemented in coordinate space
(so-called Ioffe time) distribution functions \cite{1}. In accordance with
their properties under charge conjugation one introduces these coordinate space
distributions as one-dimensional Fourier sine and cosine transforms of the
momentum space quark and gluon distributions:

\begin{eqnarray}
{\cal Q} (y^+, Q^2) & = & \int_0^1 dx \sin (\frac{My^+}{2} x) [q (x, Q^2) + 
\bar{q} (x, Q^2)],\\
{\cal Q}_{valence} (y^+, Q^2) & = & \int_0^1 dx \cos (\frac{My^+}{2} x) [q (x, 
Q^2) -
\bar{q}(x, Q^2)],\\
{\cal G}(y^+, Q^2) & = & \int_0^1 dx \cos (\frac{My^+}{2}x) x g (x, Q^2).
\end{eqnarray}

\noindent Consider first the coordinate space parton distributions of a free 
nucleon. We
start from realistic input distributions using the CTEQ4L parametrization
\cite{2} at $Q_0^2 = 1.4 \; GeV^2$, perform the QCD (DGLAP) evolution to $Q^2 =
4 \; GeV^2$ and then take the Fourier transforms (1--3) to translate the
distributions into coordinate space. The result \cite{3} is shown in
Fig. 1. Note that ${\cal Q} (y^+, Q^2)$ and ${\cal G} (y^+, Q^2)$ extend to 
distances
far beyond the diameter of the nucleon. In fact they grow continuously,
reflecting the strong rise of the structure function $F_2 (x, Q^2)$ observed at
$x < 10^{-3}$. Even the valence quark distribution has a pronounced tail
extending beyond the nucleon size. The interpretation in the lab frame is
simple: at very small Bjorken-$x$ corresponding to large longitudinal distances
        the virtual photon converts into a beam of partons which propagate
        along the light cone and interact with partons of the target nucleon,
        probing its sea quark and gluon content. This "beam" stretches over
        length scales much larger than the size of the nucleon itself. The 
partonic
        composition of that beam is dominated by a steadily growing number of 
gluons at
        larger distances.

\begin{figure}[h]
\centerline{\input{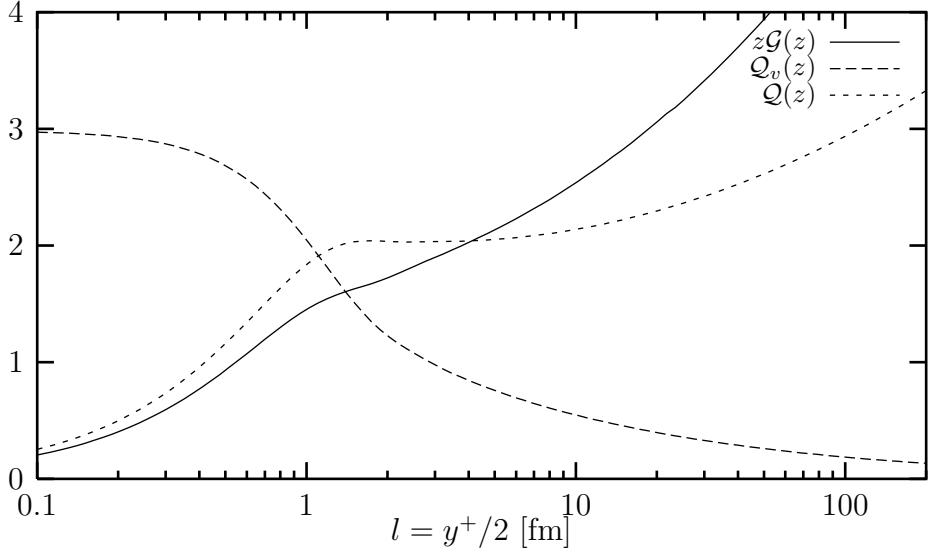}}
\caption{
Coordinate space quark and gluon distributions at $Q^2 = 4 \:
GeV^2$. Dashed curve: Valence quarks; dotted curve: total quark and antiquark
distribution; solid curve: gluons. For details see ref. \cite{3}.}
\end{figure}

It is instructive to examine parton distributions of nuclei using the same
picture. A detailed momentum space analysis of these distributions has been
performed in refs. \cite{4}, combining data from deep-inelastic lepton-nucleus
scattering and Drell-Yan lepton pair production in proton-nucleus
collisions. Based on these data the quark and gluon content of the nuclear
distributions was extracted using a DGLAP evolution analysis. One can then take
the ratio of Fourier transforms of the nuclear and free nucleon distributions:
\begin{equation}
{\cal R}(y^+, Q^2) = \frac{{\cal Q}^{A} (y^+, Q^2)}{{\cal Q}^N (y^+, Q^2)}
\end{equation}
for quark distributions, and analogous ratios for gluon distributions. Nuclear
effects can now be analysed in coordinate space. The result \cite{3} shown 
inFig.~2 for $^{40}Ca$ clearly demonstrates the most prominent features. 
Effects
of binding and Fermi motion which modify the structure functions at
longitudinal distances $l = y^+ /2$ smaller than the nucleon diameter $(l<2
fm)$ are evidently marginal. The leading effect is shadowing, the reduction of
the ratios ${\cal R}$ substantially below one, due to coherent multiple
scattering of the parton "beam" from at least two nucleons in the target
nucleus. This effect starts as soon as the propagation length of quark and
gluon fluctuations of the virtual photon exceeds the average distance
between nucleons in the nucleus $(l > d \simeq 2 fm)$. The interesting
feature is again the prominent role of gluons in this process. The measured
shadowing effect is represented by the ratio ${\cal R}_{F_2}$ of the
corresponding $F_2$ structure functions for nuclei and free nucleons. This 
ratio
reflects the shadowing effect on the sum of quark and antiquark distributions
seen directly by the virtual photon. The indirect effect of gluon shadowing,
expressed in terms of the ratio ${\cal R}_g = {\cal G}^A (y^+, Q^2)/ {\cal G}^N
(y^+, Q^2)$, is obviously very strong. While this ratio is not directly
observable, it certainly indicates a large effective cross section for gluons
interacting with nucleons.

\begin{figure}[h]
\centerline{\input{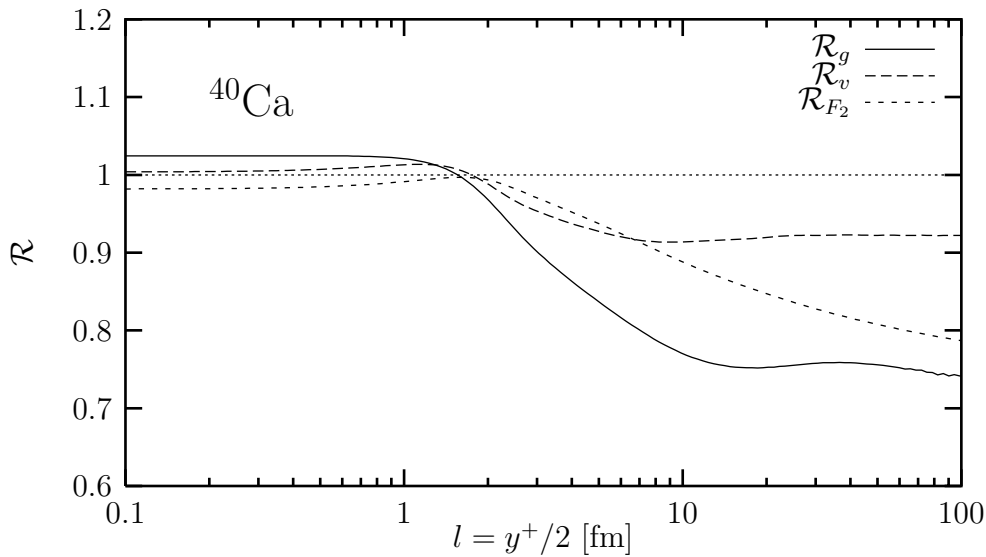}}
\caption{ Ratios of nuclear $(^{40}Ca)$ and free nucleon coordinate
space distributions for gluons (solid), valence quarks (dashed) and for the
structure functions $F_2^{A, N}$ (dotted) at $Q^2 = 4 \: GeV^2$. For further
details see ref. \cite{3}.}
\end{figure}

\section{Partons versus hadrons: generalized nucleon polarizabilities}
At intermediate $Q^2\lsim 
%\stackrel{<}{\sim} 
1 GeV^2$ a detailed analysis of virtual Compton
scattering on the nucleon should provide insights into the transition from
partonic to hadronic degrees of freedom in the nucleon. Interesting quantities
to look at in this context are the generalized ($Q^2$-dependent)
electromagnetic polarizabilities. We investigate:
\begin{enumerate}
\item[a)] the sum of electric and magnetic polarizabilities
\begin{equation}
\Sigma (Q^2) = (\alpha + \beta)_{Q^2} = \frac{1}{2 \pi^2}
\int_{\omega_0}^\infty \frac{d \omega}{\omega^2} \sigma_T (\omega, Q^2),
\end{equation}

where $\sigma_T (\omega, Q^2)$ denotes the total cross section for scattering
of a transverse virtual photon at fixed $Q^2 = \vec{q}\: ^2 - \omega^2 > 0$ 
from a
nucleon, and the integration is taken over the photon energy $\omega$ from
(pion-production) threshold to infinity. The $\omega^{-2}$ weighting in this
polarizability integral focuses on the low energy part of the excitation
spectrum, but probed over a certain range of "resolution" $Q^2$;
\item[b)] the spin polarizability
\begin{equation}
\gamma (Q^2) = \frac{1}{4 \pi^2} \int_{\omega_0}^{\infty} \frac{d
\omega}{\omega^3} [\sigma_{1/2} (\omega, Q^2) - \sigma_{3/2} (\omega, Q^2)]
\end{equation}
which involves the difference of helicity $1/2$ and $3/2$ photon-nucleon cross
sections, accessible by polarized electroproduction measurements on the nucleon
at fixed $Q^2$. This polarizability has a characteristic $\omega^{-3}$
weighting under the integral. We recall that the corresponding integral with
$\omega^{-1}$ gives the Gerasimov-Drell-Hearn sum rule.
\end{enumerate}
In the limit of large $Q^2$, the polarizabilities $\Sigma (Q^2)$ and $\gamma
(Q^2)$ turn into certain moments of the structure functions $F_1$ and $g_1$
measured in unpolarized and polarized deep-inelastic lepton-nucleon scattering:
\begin{eqnarray}
\Sigma (Q^2) \longrightarrow \frac{2e^2M}{\pi Q^4}
\int_0^1 dx \frac{x}{1-x} F_1(x),\\
\gamma (Q^2) \longrightarrow  \frac{4e^2 M^2}{\pi Q^6}
\int_0^1 dx \frac{x^2}{1-x} g_1 (x).
\end{eqnarray}

\begin{figure}[bh]
\vspace*{-1cm}
\centerline{\epsfig{figure=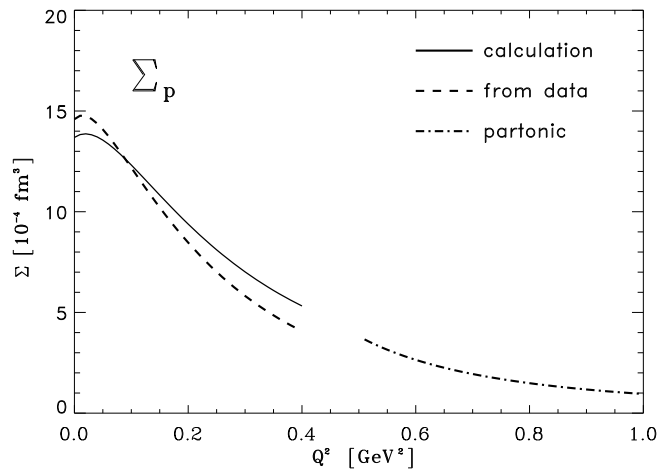,width=100mm}}
\vspace*{-1cm}
\caption{Generalized proton electromagnetic polarizability
$\Sigma_{\rho} (Q^2)$ (see text and ref. \cite{5}). The calculated full line
includes relativistic pion loops and the $\Delta (1232)$ resonance. Data
(dashed) from inelastic electron-proton scattering. Dash-dotted curve: partonic
description using the structure function $F_1$.}
\vspace*{-.3cm}
\end{figure}

The interesting question is then the following: how does the QCD description of
$\Sigma (Q^2)$ and $\gamma (Q^2)$ in terms of quarks and gluons at large $Q^2$
turn into the hadronic low $Q^2$ description of the same quantities?
At low $Q^2$, QCD with light (u-, d- and s-) quarks translates into a chiral
effective Lagrangian of Goldstone bosons (pions, kaons,...) coupled to baryons
and vector mesons. This effective field theory has been used \cite{5} at the
level of one-loop chiral perturbation theory in the pion-nucleon sector to
evaluate the nucleon polarizabilities $\Sigma (Q^2)$ and $\gamma (Q^2)$ for
$Q^2 \stackrel{<}{\sim} 0.5\: GeV^2$. This framework emphasizes the role of the
pion cloud and the $N \to \Delta$ excitation of the nucleon in response to the
electromagnetic field.
%\begin{figure}[bhH]
%\vspace*{-1cm}
%\centerline{\epsfig{figure=figure3.ps,width=100mm}}
%\vspace*{-1cm}
%\caption{Generalized proton electromagnetic polarizability
%$\Sigma_{\rho} (Q^2)$ (see text and ref. \cite{5}). The calculated full line
%includes relativistic pion loops and the $\Delta (1232)$ resonance. Data
%(dashed) from inelastic electron-proton scattering. Dash-dotted curve: partonic
%description using the structure function $F_1$.}
%\vspace*{-1cm}
%\end{figure}

At high $Q^2$, on the other hand, nucleon structure is realised in terms of 
its parton (quark and
gluon) content. We have explored whether low-$Q^2$ chiral dynamics matches
high-$Q^2$ partonic structure in $\Sigma (Q^2)$ and $\gamma (Q^2)$. The results
\cite{5} are presented in Figs. 3, 4. Let us first discuss the $Q^2$-dependent
electromagnetic polarizability $\Sigma_{p}$ of the proton (Fig. 3). The
chiral dynamics calculation (solid line) summarizes the response of the pion
cloud and through resonance (primarily $\Delta (1232)$) excitation and compares
very well with the data (dashed curve) derived from inelastic electron
scattering in the range $Q^2 \stackrel{<}{\sim} 0.5\: GeV^2$. The downward
extrapolation of the parton distribution $F_1$ from deep-inelastic scattering
(dash-dotted curve) determines $\Sigma (Q^2)$ at large $Q^2$. Both descriptions
evidently meet at intermediate $Q^2$, indicating a smooth crossover from
partonic to hadronic degrees of freedom.

%\begin{figure}[h]
%\centerline{\epsfig{figure=figure4.ps,width=100mm}}
%\caption{generalized proton spin polarizability $\gamma (Q^2)$ as
%calculated in ref. \cite{5}. Pion loop (diamagnetic) and $\Delta$ resonance
%(paramagnetic) contributions are shown separately. Their sum (solid curve) is
%compared with the partonic prediction using the spin structure function $g_1$.}
%\end{figure}

The result for the $Q^2$-dependent spin polarizability has similar features
(Fig.~4). On the low-$Q^2$ side we see the well-known balance
between the diamagnetic response of the pion cloud and the spin-paramagnetic
effect of the $N \to \Delta$ (spin $1/2$-to-spin $3/2$) transition. Dia- and
paramagnetism enter with opposite signs and comparable magnitudes, so that the
resulting $\gamma (Q^2)$ is small and changes sign at $Q^2 \simeq 0.4 \:
GeV^2$. The downward extrapolation from high $Q^2$ can be made using the spin
structure function $g_1$ from polarized deep-inelastic scattering. Again it
matches the low-$Q^2$ hadronic calculation remarkably well.

\begin{figure}[tbh]
\centerline{\epsfig{figure=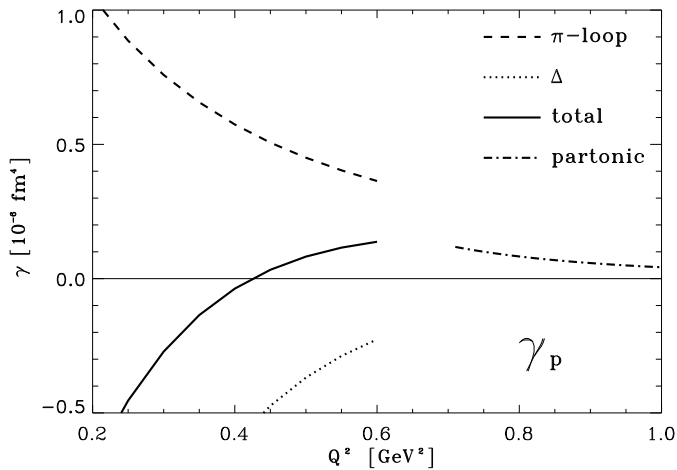,width=100mm}}
\caption{generalized proton spin polarizability $\gamma (Q^2)$ as
calculated in ref. \cite{5}. Pion loop (diamagnetic) and $\Delta$ resonance
(paramagnetic) contributions are shown separately. Their sum (solid curve) is
compared with the partonic prediction using the spin structure function $g_1$.}
\end{figure}

\section{Low-energy QCD with strange quarks: Chiral $SU(3)$ Dynamics}
QCD with massless u-, d- and s-quarks has a chiral $SU(3)_L \times SU(3)_R$
symmetry. As a consequence of strong QCD forces this symmetry is spontaneously
broken. It is also explicitly broken by the quark masses. The symmetry breaking
pattern is manifest in the low-energy hadron spectrum, with the light
pseudoscalar mesons representing the Goldstone bosons of the spontaneously
broken symmetry. The strange quark is special since its mass is intermediate
between "light" and "heavy". One of the key questions is then the following: to
what extent do the {\it symmetries} of QCD govern strong interaction {\it
dynamics}, and what is the role of the strange quark in this context?

Progress has recently been made in developing a framework to deal with these
problems. The starting point is the chiral effective Lagrangian, a theory of
Goldstone bosons (pions, kaons and eta mesons) coupled to the octet of
baryons. It is designed according to the rules set by QCD symmetries
\cite{6}. This effective Lagrangian introduces a characteristic scale, $4 \pi 
f_{\pi} \sim 1\: GeV$, where $f_{\pi} = 92.4\: MeV$ is the pion decay constant.

Spontaneously broken chiral symmetry also implies that the low-energy 
interactions
of Goldstone bosons are weak: the leading behaviour of any amplitude for
scattering of a pseudoscalar meson on a baryon goes like
\begin{equation}
T_{meson - baryon} = const \cdot \frac{E}{f_{\pi}^2} + ...,
\end{equation}

\noindent where $E$ is the meson energy in the center-of-mass system. The 
constant in eq.~(9) is specific for each meson-baryon channel and completely 
determined by
    $SU(3)$ symmetry. Pions close to threshold with $E \simeq m_{\pi}$ have
    small scattering amplitudes. In this case chiral perturbation theory 
(ChPT), the
    systematic expansion of observables in powers of energy of momentum (or
    $m_{\pi}$) is a useful concept. When strange quarks are involved, the
    scattering amplitudes scale as $E/f^2_{\pi}$ with $E \geq m_K$, and the
    driving terms in (9) become sizable. One therefore expects that ChPT is
    only of limited value once strangeness is included. A way to proceed which
    proves to be quite successful is chiral $SU(3)$ dynamics \cite{7}: a
    non-perturbative coupled-channel approach based on the chiral $SU(3) \times
    SU(3)$ effective meson-baryon Lagrangian. The basic strategy is first to 
generate
    Born terms $T^{(0)}_{ij}$ of the multi-channel meson-baryon T-matrix from
    the chiral effective Lagrangian and then to perform a partial loop 
summation to all orders using a
                         Lippmann-Schwinger equation:

\begin{equation}
T = [1 - T^{(0)} \cdot G]^{-1} T^{(0)},
\end{equation}
\noindent with an appropriate Green function $G$. Introducing a limited number 
of finite-range
parameters, a remarkably good description of a large amount of cross section
can be achieved. This method has been successfully applied \cite{7} to $KN$
scattering, the coupled $ \{\bar{K}N, \pi Y, \eta Y\}$ and $\{ \pi N, KY, \eta 
N
\}$ multi-channel systems, and to the photoproduction of $\eta$ and $K$
mesons. Simplified but otherwise very similar calculations are reported
elsewhere at this conference \cite{8}.

While all previous calculations have dealt with $s$-wave dynamics, the present
focus is on the systematic incorporation of $p$-waves as required by a
multitude of measured meson-baryon angular distributions and polarization
observables. We have now reached the stage where the complete $s$- and $p$-wave
chiral $SU(3)$ dynamics is well under control \cite{9}. As an example Fig.~5
shows predicted cross sections for kaon photoproduction once a limited set of
parameters has been fixed to reproduce a large variety of $\pi N \to \eta N$, 
$K
\Lambda$ and $K \Sigma$ cross sections and angular distributions. The $l = 1$
partial waves are evidently important in such channels immediately above
threshold. With the inclusion of $p$-waves, further detailed tests of the 
chiral
$SU(3)$ effective Lagrangian are now possible.

A forthcoming necessary step is to incorporate
the axial $U (1)$ anomaly and the dynamics of the $\eta '$. This is where
the gluonic sector of QCD should have its impact.

\begin{figure}[h]
\centerline{\epsfig{figure=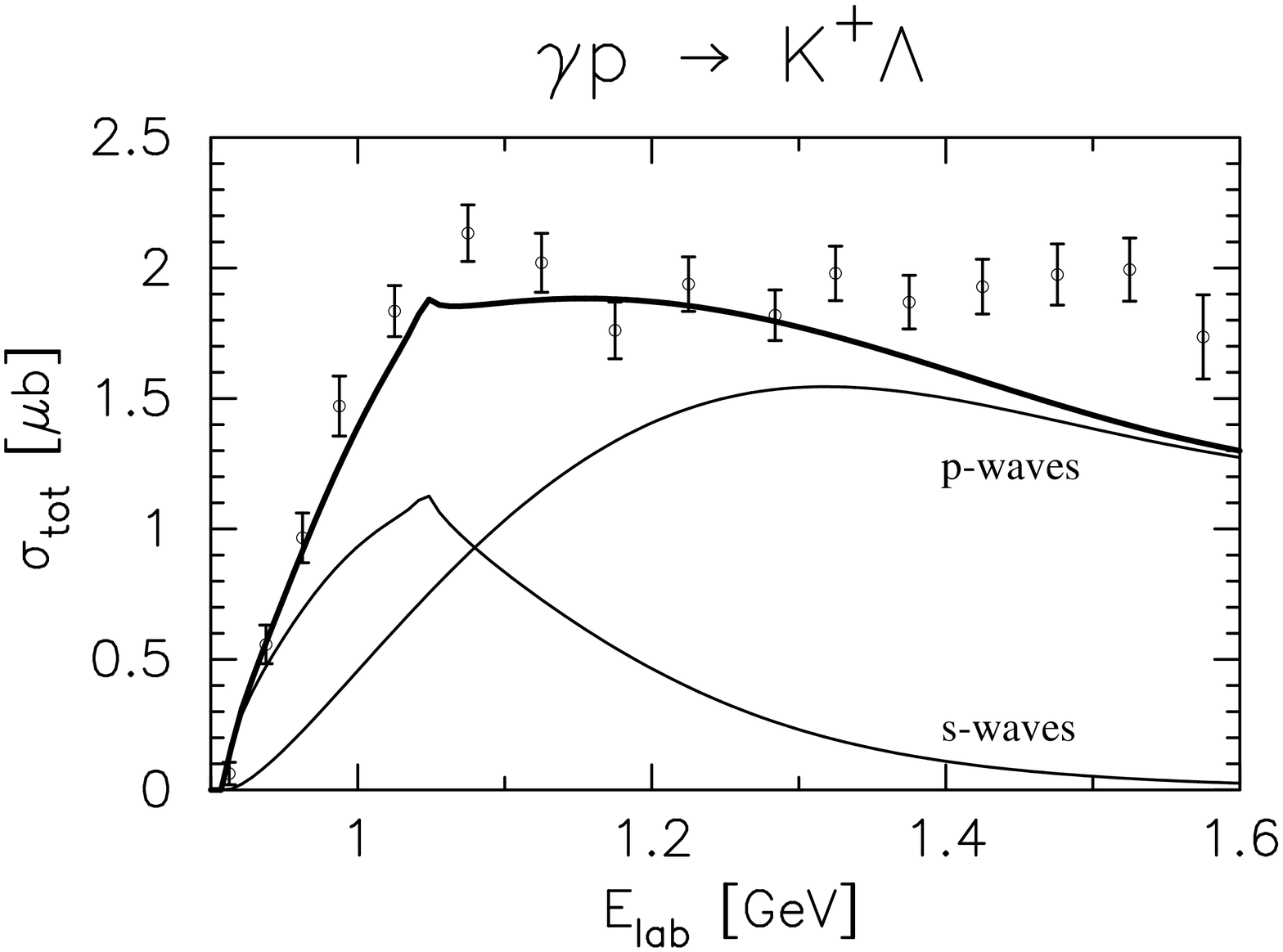,width=50mm,angle=-90}
\epsfig{figure=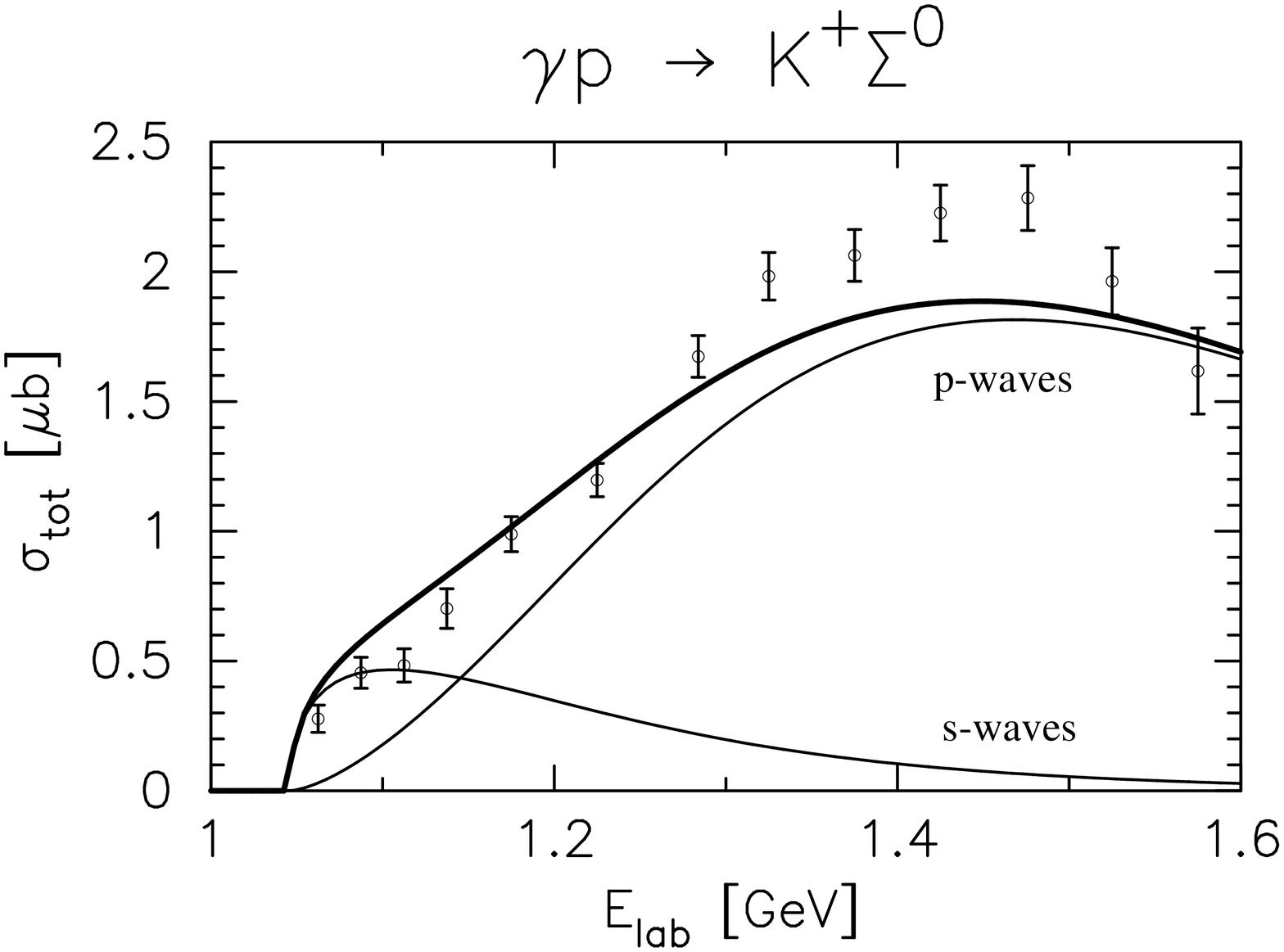,width=50mm,angle=-90}}
\caption{Kaon photoproduction cross sections using s- and p-wave chiral
$SU(3)$ dynamics \cite{9}, a non-perturbative coupled channel approach based on
the chiral meson-baryon Lagrangian \cite{7}. The data are taken from ref. 
\cite{10}.}
\end{figure}

\section{QCD Sum Rules: some recent developments}
The QCD sum rule method of treating the non-perturbative dynamics of QCD was
developed twenty years ago (SVZ sum rules \cite{11}). It connects the expansion
of a correlation function in terms of vacuum condensates (the operator product
expansion) with the spectrum of this correlation function via dispersion 
relations. These sum rules have been applied to understand the masses and 
properties of a
variety of hadrons emphasizing their role as excitations of the condensed QCD 
vacuum.

QCD sum rules have also been used in more recent times to arrive at estimates
for possible in-medium mass shifts of vector mesons \cite{12}. The validity of
such estimates has been under debate, however, for several
reasons. Uncertainties exist at the level of factorization assumptions used to
approximate four-quark condensates $\langle \bar{q} \Gamma q \bar{q} \Gamma q 
\rangle$
in terms of $\langle \bar{q} q \rangle^2$, the square of the standard chiral
condensate.  Furthermore, for broad structures such as the $\rho$ meson, with 
its
large vacuum decay width further magnified by in-medium reactions, the QCD sum
rule analysis does not provide a reliable framework to extract a "mass shift"
in medium. The situation is more comfortable for the $\omega$ meson which has a
vacuum width twenty times smaller than that of the $\rho$ meson and may
have a much better chance to survive as a quasi-particle in nuclear matter
\cite{13}.

We have recently re-examined these questions \cite{14} in search for
model-independent sum-rule constraints which do not suffer from the
uncertainties introduced by four-quark condensates. We exemplify these 
constraints for the case of the $\omega$ meson spectral distribution and its
changes in the nuclear medium.

The starting point is the current-current correlation function
\begin{equation}
\Pi_{\mu \nu} (q) = i \int d^4 x e^{iq \cdot x} \langle {\cal T}\: j_{\mu} (x)
j_{\nu} (0) \rangle = \left( \frac{q_{\mu} q_{\nu}}{q^2} - g_{\mu \nu}
\right) \Pi (q^2),
\end{equation}
and we work with the spectrum
\begin{equation}
R (s) = \frac{12 \pi}{s} Im \Pi (s).
\end{equation}

\noindent In vacuum $R (s)$ is directly related to $\sigma (e^+ e^- \to 
hadrons)$. In
dense and hot hadronic matter it enters into the analysis of lepton pair
production in high-energy heavy-ion collisions.

One proceeds now as follows. First $\Pi (q^2)$ is expanded at large spacelike
$q^2$ (i. e. for $Q^2 = -q^2$ positive and large) in powers of $1/Q^2$ using
the QCD operator product expansion. At the same time $\Pi (q^2)$ is written in
the form of a dispersion relation. Then a Borel transformation is performed
which effectively reduces the weight on the uncertain high-energy parts of the
spectrum. Let us focus on the $\omega$ meson with the isoscalar current 
$j_{\mu} =
\frac{1}{6} (\bar{u} \gamma_{\mu} u + \bar{d} \gamma_{\mu} d)$ and discuss the
vacuum case first. One finds

\begin{equation}
\int^{\infty}_{0} ds\: R(s)e^{-s / {\cal M}^2} = \frac{1}{6} (1 +
\frac{\alpha_s}{\pi}) {\cal M}^2 + \frac{C}{{\cal M}^2} +...
\end{equation}

\noindent where the Borel mass ${\cal M}$ is a technical scale parameter and 
$C$ is a
combination of quark and gluon condensates:

\begin{equation}
C = \frac{2 \pi^2}{3} \left( \langle m_u \bar{u} u + m_d \bar{d} d \rangle +
\frac{1}{12} \langle \frac{\alpha_s}{\pi} G_{\mu \nu} G^{\mu \nu} \rangle
\right) .
\end{equation}

\noindent A negligibly small quark mass term has been dropped on the r.h.s. of 
eq. (13). Terms of
order ${\cal M}^{-4}$ involve combinations of (uncertain) four-quark
condensates. In a nuclear medium the condensate term (14) receives
density-dependent corrections, the leading one being proportional to the first
moment of the quark distribution in the nucleon.

The vacuum spectrum of the $\omega$ meson has the characteristic behaviour
shown in Fig. 6: a resonant part below a scale $s_0$ followed by a continuum
$R_c (s)$ which approaches the perturbative QCD limit for $s > s_0$:

\begin{equation}
R_c \simeq \frac{1}{6} (1 + \frac{\alpha_s}{\pi}) \Theta (s - s_0).
\end{equation}

\noindent Splitting the spectrum into resonance and continuum and choosing
${\cal M} > \sqrt{s_0}$, a term-by-term comparison in eq. (13) gives a set of
sum rules for the moments of $R(s)$. The lowest ones (in vacuum) are

\begin{eqnarray}
\int^{s_0}_0 ds \: R(s) & = & \frac{s_0}{6} (1 + \frac{\alpha_s}{\pi}),\\
\int^{s_0}_0 ds \: s \:R(s) & = & \frac{s^2_0}{12} (1 + \frac{\alpha_s}{\pi})- 
C,
\end{eqnarray}

\noindent with the condensate term $C$ of eq. (14). Higher moments of $R(s)$
successively introduce condensates of higher dimensions.

\begin{figure}[h]
\centerline{\epsfig{figure=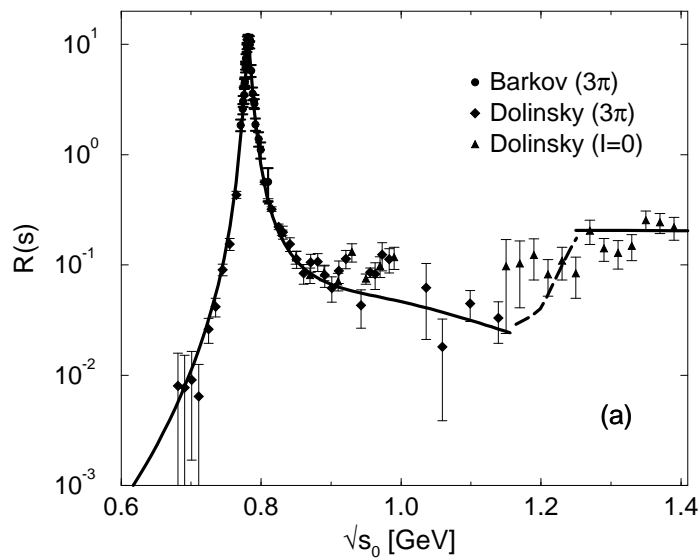,width=100mm}}
\caption{Spectrum R(s) in the $\omega$ meson channel as calculated in
ref. \cite{13,14}. The data points refer to $e^+ e^- \to 3 \pi$ and $e^+ e^-
\to$ hadrons $(I = 0)$.}
\end{figure}

Let us now make an attempt to unify QCD sum rules with chiral symmetry and 
current
algebra. Recall that the scale for spontaneous chiral symmetry breaking,
$\Delta = 4 \pi\; f_{\pi} \sim 1\: GeV$, is realized as a characteristic gap in
the low-mass hadron spectrum. The light vector mesons are the lowest $q 
\bar{q}$
dipole $(J^{\pi} = 1^-)$ excitations of the QCD vacuum, with their masses
located just under the gap $\Delta$.

We propose \cite{14} that the scale $\sqrt{s_0}$ which separates the
hadronic (resonance) sector from the quark-antiquark continuum in the QCD sum
rule analysis, should be identified with the gap $\Delta$, setting

\begin{equation}
\sqrt{s_0} = \Delta = 4 \pi \;f_{\pi}.
\end{equation}

\noindent That this makes some sense can be seen instantly by returning to the
Vector Meson Dominance (VMD) model for the resonant part of $R (s)$. In this
model, taking the zero width limit, we have

\begin{equation}
R (s) = \frac{4 \pi^2}{3} \frac{m^2_{\omega}}{g^2} \delta (s-m^2_{\omega}) +
R_c (s)
\end{equation}

\noindent with the vector coupling constant $g \simeq 6$. From the sum rule 
(16)
one finds

\begin{equation}
\frac{8 \pi^2}{g^2} \frac{m^2_{\omega}}{s_0} = 1 + \frac{\alpha_s}{\pi}.
\end{equation}

\noindent In fact this equation holds both for $\omega$ and $\rho$ meson in the
zero width limit, with degenerate masses $m_V = m_{\rho} =
m_{\omega}$. Inserting $s_0 = 16 \pi^2 f_{\pi}^2$ one immediately recovers the
celebrated current algebra (KSFR) relation

\begin{equation}
m_V = \sqrt{2} g f_{\pi} ,
\end{equation}

\noindent apart from a small perturbative QCD correction. Note that the second
sum rule (17) gives the interesting further constraint $g = 2 \pi$ (up to small
$\alpha_s$ and condensate corrections which drive $g$ closer to its empirical 
value).

The in-medium downward shift of the vector meson mass in the zero-width limit
found in ref. \cite{12} was about 15 \% at nuclear matter density, $\rho =
\rho_0 = 0.17 \: fm^{-3}$, and came primarily through the downward shift of 
$s_0$ in the
QCD sum rule analysis. With the identification (18) in the dipole sum rules
(16, 17), this can now be easily understood in terms of the in-medium reduction
of the gap $\Delta (\rho) = 4 \pi \: f_{\pi} (\rho) = \sqrt{s_0 (\rho)}$. The
pion decay constant (in fact, the one related to the time component of the
axial current in hadronic matter) is proportional to the square root of the
chiral condensate $\langle \bar{q} q \rangle$. Its leading dependence on baryon
density is controlled by the nucleon sigma term $\langle N |m_q \bar{q} q| N
\rangle$ which induces an approximate 30 \% reduction of the magnitude of 
$\langle \bar{q} q \rangle$ at
$\rho = \rho_0$.

While such relationships are obscured for the $\rho $ meson by its very
large in-medium width, they may well be realized for the much narrower 
$\omega$ meson. In
fact explicit calculations \cite{13} of the in-medium $\omega$ meson  spectrum
using the chiral $SU(3) \times SU(3)$ Lagrangian with inclusion of anomalous 
couplings
from the Wess-Zumino action, turn out to be fully consistent with the QCD sum
rule analysis. The suggested in-medium mass shift may even lead to
nuclear bound states of $\omega$ mesons \cite{15,16}, an exciting perspective.


\begin{thebibliography}{99}
\bibitem{1} P.~Hoyer and M.~V\"anttinen, Z.~Phys.~{\bf C 74} (1997) 113;
B.~L.~Ioffe,  Phys.~Lett.~ {\bf B 30} (1969) 123. 
\bibitem{2} H.~L.~Lai et al., Phys.~Rev.~{\bf D 55} (1997) 1280
\bibitem{3} M.~V\"anttinen, G.~Piller, L.~Mankiewicz, W.~Weise and
K.~J.~Eskola, Eur.~ Phys.~J.~{\bf A 3} (1998) 351.
\bibitem{4} L.~Frankfurt, M.~Strikman and S.~Liuti, Phys.~Rev.~Lett.~{\bf 65}
(1990) 1725; K.~J.~Eskola, Nucl.~Phys.~{\bf B 400} (1993) 240; K.~J.~Eskola,
V.~J.~Kolhinen and P.~V.~Ruuskanen, hep-ph/9802350, Nucl.~Phys.~B (to appear).
\bibitem{5} J.~Edelmann, N.~Kaiser, G.~Piller and W.~Weise, Nucl.~Phys.~{\bf A
641} (1998) 119.
\bibitem{6} see e. g.: V.~Bernard, N.~Kaiser and U.-G.~Mei\ss ner,
Int.~J.~Mod.~Phys.~{\bf E 4} (1995) 193.
\bibitem{7} N.~Kaiser, P.~B.~Siegel and W.~Weise, Nucl.~Phys.~{\bf A 594}
(1995) 325; N.~Kaiser, T.~Waas and W.~Weise, Nucl.~Phys.~{\bf A 612} (1997)
297.
\bibitem{8} E.~Oset, this conference
\bibitem{9} J.~Caro~Ram\'on, N.~Kaiser, S.~Wetzel and W.~Weise, preprint 
(1999),
to be published.
\bibitem{10} M.~Q.~Tran et al., Phys.~Lett.~{\bf B 445} (1998) 20.
\bibitem{11} M.~Shifman, A.~Vainshtein and V.~Zakharov, Nucl.~Phys.~{\bf B 147}
(1979) 385; 448.
\bibitem{12} T.~Hatsuda and S.~H.~Lee, Phys.~Rev.~{\bf C 46} (1992) R 34.
\bibitem{13} F.~Klingl, N.~Kaiser and W.~Weise, Nucl.~Phys.~{\bf A 624} (1997)
527.
\bibitem{14} F.~Klingl and W.~Weise, Eur.~Phys.~J.~A (1999), in print.
\bibitem{15} K.~Tsushima, D.~L.~Lu, A.~W.~Thomas and K.~Saito, Phys.~Lett.~{\bf
B 43} (1998) 26.
\bibitem{16} F.~Klingl, T.~Waas and W.~Weise, Nucl.~Phys.~A (1999), in print.
\end{thebibliography}
\end{document}